\documentclass[3p,times]{elsarticle}
\usepackage{amssymb}
\usepackage{epsfig}
\usepackage{mathrsfs}
\usepackage{graphicx}
\usepackage{amsmath}
\usepackage{dcolumn}
\usepackage{bm}
\usepackage{longtable}
\usepackage{threeparttablex}
\usepackage{multirow}
\usepackage{booktabs}
\usepackage{color}
\usepackage{pdflscape}
\usepackage[figuresright]{rotating}
\usepackage[version=4]{mhchem}
\usepackage[labelfont=bf,singlelinecheck=false,font=footnotesize]{caption}
\usepackage{hyperref}
\usepackage{array}
\newcolumntype{C}[1]{>{\centering\arraybackslash}p{#1}}
\biboptions{square,sort&compress}

\journal{Chemical Physics Letters}

\begin{document}
\begin{frontmatter}
\title{
Double Electron Attachment and Double Ionization Potential Equation-of-Motion Coupled-Cluster Approaches with Full and Active-Space Treatments of 4-Particle--2-Hole and 4-Hole--2-Particle Excitations and Three-Body Clusters
}
\author[label1]{Jun Shen}
\author[label1]{Karthik Gururangan}
\author[label1,label2]{Piotr Piecuch\corref{cor1}}
\address[label1]{Department of Chemistry, Michigan State University, East Lansing, Michigan 48824, USA}
\address[label2]{Department of Physics and Astronomy, Michigan State University, East Lansing, Michigan 48824, USA}
\ead{piecuch@chemistry.msu.edu}
\cortext[cor1]{Corresponding author}
\begin{abstract}
The double electron attachment (DEA) and double ionization potential (DIP) equation-of-motion coupled-cluster (EOMCC) methods including up to 4-particle--2-hole (4$p$-2$h$) and 4-hole--2-particle (4$h$-2$p$) excitations on top of coupled-cluster singles, doubles, and triples (CCSDT), denoted DEA-EOMCCSDT(4$p$-2$h$) and DIP-EOMCCSDT(4$h$-2$p$), have been
efficiently implemented in full and active-space forms. 
The resulting methods are applied to determine the ground and low-lying excited states of methylene, the singlet--triplet gap of trimethylenemethane, and the lowest singlet and triplet DIPs of 23 atoms 
and molecules. 
In all cases considered, the active-space DEA/DIP-EOMCC approaches recover the highly accurate parent
DEA-EOMCCSDT(4$p$-2$h$)/DIP-EOMCCSDT(4$h$-2$p$) data at small fractions of the computational costs.
\end{abstract}


%
\end{frontmatter}

\section{Introduction}
\label{sec1}
The single-reference coupled-cluster (CC) theory \cite{cizek1,cizek4} and its 
equation-of-motion (EOM) \cite{eomcc3} extension to electronically excited and electron attached and ionized states are among
the most effective tools for high-accuracy quantum-chemical
calculations. In this work, we focus on the double electron attachment (DEA) and double ionization potential (DIP) variants of EOMCC \cite{dipea1,dipea3,dip-stanton,dipea5,dipea6,kus-krylov-2011,kus-krylov-2012,jspp-dea-dip-2013,jspp-dea-dip-2014,jspp-dea-dip-2017,jspp-dea-dip-2021,dipea7,DIP-EOMCCSDT,deprince-dipeomccsdt,krylov-dipea-eomccsdt}, which provide transparent and rigorously spin-adapted frameworks for treating certain types of multireference (MR) problems, including biradical species and single-bond dissociation,
through double electron attachment to or removal from an underlying closed-shell 
$N$-electron reference species. These approaches can also directly
compute the energies of doubly attached and doubly ionized states, the latter being especially relevant to Auger electron spectroscopy \cite{ghosh-auger,krylov-auger-1,krylov-auger-2,auger-benzene} and studies of
strong-field-induced chemical reactivity \cite{h3plus}.

The standard DEA/DIP-EOMCC approximations, namely DEA-EOMCCSD(3$p$-1$h$)
and DIP-EOMCCSD(3$h$-1$p$), incorporate the 2-particle/2-hole 
(2$p$/2$h$) and 
3-particle--1-hole/3-hole--1-particle (3$p$-1$h$/3$h$-1$p$)
correlations on top of the 
$N$-electron CC singles and doubles (CCSD) \cite{ccsd,ccsd2,ccsdfritz,osaccsd} state. Although computationally practical, these methods work well
only for states dominated by 2$p$/2$h$ 
excitations. Unfortunately, they often fail when 3$p$-1$h$/3$h$-1$p$ effects become 
important, as is the case of low-lying states of biradicals and 
in stretched regions of bond-breaking potentials. In
such cases, inclusion of the 4-particle--2-hole/4-hole--2-particle
(4$p$-2$h$/4$h$-2$p$) excitations, in addition to the 2$p$/2$h$ and 3$p$-1$h$/3$h$-1$p$ terms on top of CCSD, is required to obtain quantitative 
accuracy, as in the DEA-EOMCCSD(4$p$-2$h$) and DIP-EOMCCSD(4$h$-2$p$) methodologies 
\cite{jspp-dea-dip-2013, jspp-dea-dip-2014, jspp-dea-dip-2017, jspp-dea-dip-2021}. 
More recently, we have implemented the higher-level 
DIP-EOMCCSDT(4$h$-2$p$)
\cite{DIP-EOMCCSDT}
approach, which treats the 2$h$, 3$h$-1$p$, and 4$h$-2$p$
excitations on top of the $N$-electron CC singles, doubles, and triples (CCSDT) \cite{ccfullt,ccfullt2} state  and demonstrated that the inclusion of 
three-body clusters is essential for obtaining accurate DIPs
by providing a balanced description of the 
$N$- and $(N-2)$-electron species (cf., also, Ref.\ \cite{deprince-dipeomccsdt}).
While preparing this article, we became aware of an independent 
implementation of the full 
DIP-EOMCCSDT(4$h$-2$p$) method as well as its
DEA-EOMCCSDT(4$p$-2$h$) companion in Ref.\ \cite{krylov-dipea-eomccsdt}. 
However, as elaborated on below, the emphasis of this
article is different. Rather than focusing on the 
full formulations, which
have computational steps that scale as 
$\mathscr{N}^8$ with
system size $\mathscr{N}$ and are, therefore, usually prohibitively expensive, we seek robust and computationally practical
approximations to DEA-EOMCCSDT(4$p$-2$h$) and 
DIP-EOMCCSDT(4$h$-2$p$) based on the 
ideas of 
active-space CC formalism \cite{piecuch-qtp}.

The central idea of our approach is to use active orbitals to
retain only the dominant 4$p$-2$h$/4$h$-2$p$ excitations and 
combine them with a full or active-space treatment of three-body
clusters in the underlying $N$-electron problem. This strategy can
be viewed as an extension of the previously developed 
DEA-EOMCCSD(4$p$-2$h$)\{$N_u$\} and 
DIP-EOMCCSD(4$h$-2$p$)\{$N_o$\} methods 
\cite{jspp-dea-dip-2013, jspp-dea-dip-2014, jspp-dea-dip-2017, jspp-dea-dip-2021}, 
which include all 2$p$/2$h$ and 3$p$-1$h$/3$h$-1$p$ excitations, 
along with the small subsets of 4$p$-2$h$/4$h$-2$p$ excitations 
selected by active orbitals, on top of CCSD. Here, $N_u$ and $N_o$ 
denote the numbers of active unoccupied and active occupied 
orbitals used to select the dominant 4$p$-2$h$ and 4$h$-2$p$ 
contributions. As previously shown
\cite{jspp-dea-dip-2013, jspp-dea-dip-2014, jspp-dea-dip-2017, 
jspp-dea-dip-2021}, the active-space DEA-EOMCCSD(4$p$-2$h$)\{$N_{u}$\} and DIP-EOMCCSD(4$h$-2$p$)\{$N_{o}$\} methods
are capable of recovering the results obtained with their 
DEA-EOMCCSD(4$p$-2$h$) and DIP-EOMCCSD(4$h$-2$p$) parents at small 
fractions of the computational costs. In particular, they can be 
implemented with computational steps that
scale as $\mathscr{N}^6$, with a prefactor proportional to the square of the
number of active unoccupied (DEA) or active occupied (DIP) orbitals,
making them substantially more practical than the $\mathscr{N}^8$ 
scalings of their full parents. Beyond reducing 
computational costs, active-orbital-based truncations offer a more
robust alternative to traditional perturbative treatments of 
4$p$-2$h$/4$h$-2$p$ excitations and three-body clusters used in 
DEA/DIP-EOMCC calculations, such as the
DIP-EOMCCSD(T)(a)(4$h$-2$p$) \cite{DIP-EOMCCSDT} 
and DIP-EOMCCSD(T)($\tilde{a}$)(4$h$-2$p$) \cite{4c-DIP-EOMCCSDT}
approximations to DIP-EOMCCSDT(4$h$-2$p$),
which may become unreliable in the presence of electronic
quasi-degeneracies
characteristic of more MR problems.

In this Letter, we introduce, implement, and test a family of 
active-orbital-based approximations to 
\linebreak 
DEA-EOMCCSDT(4$p$-2$h$) 
and DIP-EOMCCSDT(4$h$-2$p$), in which all 2$p$/2$h$, all 
3$p$-1$h$/3$h$-1$p$, and subsets of leading 4$p$-2$h$/4$h$-2$p$ 
excitations selected with the help of active 
orbitals are included on top of CCSDT or its active-space CCSDt 
approximation \cite{piecuch-qtp,semi0b,semi2,ccsdtq3,semi4}. The resulting
active-space DEA/DIP-EOMCC methods based on the $N$-electron 
CCSDT state are 
denoted as
DEA-EOMCCSDT(4$p$-2$h$)\{$N_u$\} and DIP-EOMCCSDT(4$h$-2$p$)\{$N_o$\}, 
whereas their counterparts based on CCSDt are designated as
DEA-EOMCCSDt(4$p$-2$h$)\{$N_u$\} and DIP-EOMCCSDt(4$h$-2$p$)\{$N_o$\}, respectively. 
We apply this hierarchy of full and active-space 
DEA/DIP-EOMCC approaches, including 
\linebreak
DEA-EOMCCSD(3$p$-1$h$),
DEA-EOMCCSD(4$p$-2$h$)\{$N_u$\},
DEA-EOMCCSD(4$p$-2$h$), DEA-EOMCCSDt(4$p$-2$h$)\{$N_u$\}, 
DEA-EOMCCSDT(4$p$-2$h$)\{$N_u$\}, DEA-EOMCCSDT(4$p$-2$h$),
DIP-EOMCCSD(3$h$-1$p$),
DIP-EOMCCSD(4$h$-2$p$)\{$N_o$\}, DIP-EOMCCSD(4$h$-2$p$), DIP-EOMCCSDt(4$h$-2$p$)\{$N_o$\}, DIP-EOMCCSDT(4$h$-2$p$)\{$N_o$\}, and DIP-EOMCCSDT(4$h$-2$p$), to study the ground and low-lying excited states of methylene and the singlet--triplet gap in trimethylenemethane (TMM), comparing the results with experiment and exact, full configuration interaction (CI) benchmarks. In addition, we use the DIP-EOMCCSD(3$h$-1$p$), DIP-EOMCCSD(4$h$-2$p$), DIP-EOMCCSD(T)(a)(4$h$-2$p$), DIP-EOMCCSDt(4$h$-2$p$)\{$N_o$\}, and DIP-EOMCCSDT(4$h$-2$p$) methods to compute the lowest singlet and triplet DIPs of 23 atoms and molecules studied in Ref.\ \cite{loos-ppbse}. In this case, we compare
the results with the near-exact DIPs obtained in Ref.\ \cite{loos-ppbse} using the CI with perturbative selection made iteratively (CIPSI) calculations \cite{sci_3,cipsi_1,cipsi_2} extrapolated to the full CI limit.
\section{Theory}
\label{sec2}
Within the DEA/DIP-EOMCC formalisms, the ground ($\mu=0$) and excited 
($\mu>0$) states of the $(N\pm2)$-electron target species are represented as
\begin{equation}
|\Psi_{\mu}^{(N\pm2)}\rangle = R_{\mu}^{(\pm2)} |\Psi_{0}^{(N)}\rangle,
\label{deadip-ansatz}
\end{equation}
where
\begin{equation}
|\Psi_{0}^{(N)}\rangle = e^{T}|\Phi\rangle
\label{eq2}
\end{equation}
is the CC ground state of the underlying $N$-electron system and 
$|\Phi\rangle$ is the corresponding reference determinant, which serves as
a Fermi vacuum. 
The cluster operator $T$ is defined as
\begin{equation}
T = \sum_{n=1}^{M_{T}} T_{n},
\label{tdef}
\end{equation}
where $M_T \le N$, with its $n$-body components given by
\begin{equation}
T_n = \sum_{i_{1} < \cdots < i_{n},\,a_{1} < \cdots < a_{n}}
t_{a_{1}\ldots a_{n}}^{i_{1} \ldots i_{n}} \,
a^{a_{1}} \cdots a^{a_{n}} a_{i_{n}} \cdots a_{i_{1}}.
\label{tamp}
\end{equation}
The double electron attachment [$R_{\mu}^{(+2)}$] and double ionization 
[$R_{\mu}^{(-2)}$] EOM operators are expanded as
\begin{equation}
R_{\mu}^{(+2)}=\sum_{n=0}^{M_{R}}R_{\mu,(n+2)p\mbox{-}nh},
\label{rnplus2}
\end{equation}
where $M_{R} \leq N$, and
\begin{equation}
R_{\mu}^{(-2)}=\sum_{n=0}^{M_{R}}R_{\mu,(n+2)h\mbox{-}np},
\label{rnminus2}
\end{equation}
with $M_{R} \leq N-2$. The many-body components of the $R_{\mu}^{(+2)}$
and $R_{\mu}^{(-2)}$ operators are given by
\begin{equation}
R_{\mu,(n+2)p\mbox{-}nh} =
\sum_{
k_{1} < \cdots < k_{n},\, a < b < c_{1} < \cdots < c_{n}
}
r^{\;\;\; k_{1} \ldots k_{n}}_{a b c_{1} \ldots c_{n}}(\mu)
\:\: a^{a} a^{b} a^{c_{1}} \cdots a^{c_{n}} a_{k_{n}} \cdots a_{k_{1}}
\label{r-nplus2-comp}
\end{equation}
and
\begin{equation}
R_{\mu,(n+2)h\mbox{-}np} =
\sum_{
k_{1} < \cdots < k_{n} < i < j,\, c_{1} < \cdots < c_{n}
}
r^{i j k_{1} \ldots k_{n}}_{\;\;\; c_{1} \ldots c_{n}}(\mu)
\:\: a^{c_{1}} \cdots a^{c_{n}} a_{k_{n}} \cdots a_{k_{1}} a_{j} a_{i},
\label{r-nminus2-comp}
\end{equation}
respectively.
Throughout this article, indices $i,j,k,l,\ldots$ ($a,b,c,d,\ldots$) are used to denote spin-orbitals occupied
(unoccupied) in the $N$-electron reference determinant $|\Phi\rangle$, while $a^{p}$ ($a_{p}$) represents the usual Fermionic creation (annihilation) operator for spin-orbital $|p\rangle$. Once the truncation levels
$M_T$ and $M_R$ are specified, the amplitudes
$r^{\;\;\; k_{1} \ldots k_{n}}_{a b c_{1} \ldots c_{n}}(\mu)$ and
$r^{i j k_{1} \ldots k_{n}}_{\;\;\; c_{1} \ldots c_{n}}(\mu)$, together with 
the corresponding vertical DEA/DIP energies $\omega_{\mu}^{(N\pm2)} = E_{\mu}^{(N\pm2)} - E_{0}^{(N)}$ associated with the
$(N\pm2)$-electron target states, are obtained by solving the
non-Hermitian eigenvalue problem
\begin{equation}
    (\overline{H}_{N,\text{open}} R_{\mu}^{(\pm 2)})_C|\Phi\rangle = \omega_{\mu}^{(\pm 2)}R_{\mu}^{(\pm 2)}|\Phi\rangle,
    \label{eomproblem}
\end{equation}
where
\begin{equation}
\overline{H}_{N} = e^{-T} H_{N} e^{T} = (H_{N} e^{T})_C
\end{equation}
is the similarity-transformed Hamiltonian corresponding to the 
$N$-electron ground-state CC problem, with 
$\overline{H}_{N,\mathrm{open}}$ representing
the diagrams of $\overline{H}_{N}$ containing external Fermion lines, and
subscript $C$ refers to the connected part of the operator product.
The eigenvalue equation
described by Eq.\ (\ref{eomproblem}) amounts to diagonalizing
the similarity-transformed Hamiltonian in the appropriate 
$(N\pm2)$-electron sector of the Fock space spanned by the set of
($n+2$)$p$-$nh$/($n+2$)$h$-$np$ excitations included in $R_{\mu}^{(\pm2)}$.

Different choices of $M_{T}$ and $M_{R}$ define the standard hierarchy of 
DEA/DIP-EOMCC approximations. 
The basic DEA-EOMCCSD(3$p$-1$h$) and DIP-EOMCCSD(3$h$-1$p$) methods 
correspond to $M_{T}=2$ and $M_{R}=1$, incorporating the 2$p$/2$h$ and 
3$p$-1$h$/3$h$-1$p$ excitations on top of CCSD. Their higher-level 
DEA-EOMCCSD(4$p$-2$h$) and DIP-EOMCCSD(4$h$-2$p$) counterparts 
\cite{jspp-dea-dip-2013, jspp-dea-dip-2014, jspp-dea-dip-2017, jspp-dea-dip-2021} 
employ $M_{T}=2$ and $M_{R}=2$, thereby extending the diagonalization
space to include the 4$p$-2$h$/4$h$-2$p$ determinants. 
In the present work, we 
focus on the more advanced DEA-EOMCCSDT(4$p$-2$h$) and DIP-EOMCCSDT(4$h$-2$p$) 
approximations defined by $M_{T}=3$ and $M_{R}=2$, which treat up to 
4$p$-2$h$/4$h$-2$p$ excitations on top of the higher-level
CCSDT description 
of the underlying $N$-electron system. The working equations for the full
DEA-EOMCCSDT(4$p$-2$h$) diagonalization problem corresponding to projections
onto the 2$p$ ($|\Phi^{ab}\rangle$),
\linebreak
3$p$-1$h$ 
($|\Phi_{\phantom{ij}k}^{abc}\rangle$), and 4$p$-2$h$ 
($|\Phi_{\phantom{ij}kl}^{abcd}\rangle$) determinants 
are given by (cf., also, Ref.\ \cite{krylov-dipea-eomccsdt})
\begin{eqnarray}
\langle \Phi^{ab} | (\overline{H}_{N,\mathrm{open}}^{(\text{CCSDT})} R_{\mu}^{(+2)})_C | \Phi\rangle
= \mathscr{A}_{ab}[
 \bar{h}_{a}^{e}r_{eb}^{}(\mu)
+\tfrac{1}{4}\bar{h}_{ab}^{ef}r_{ef}^{}(\mu)
+\tfrac{1}{2}\bar{h}_{m}^{e}r_{abe}^{\phantom{ij}m}(\mu)
+\tfrac{1}{2}\bar{h}_{an}^{ef}r_{ebf}^{\phantom{ij}n}(\mu)
+\tfrac{1}{8}\bar{h}_{mn}^{ef}r_{abef}^{\phantom{ij}mn}(\mu)
],
\label{eq2p}
\end{eqnarray}
\begin{align}
\langle \Phi_{\phantom{ij}k}^{abc} | (\overline{H}_{N,\mathrm{open}}^{(\text{CCSDT})} R_{\mu}^{(+2)})_C | \Phi\rangle
=&\mathscr{A}_{abc}[
-\tfrac{1}{2} {I^{\prime}}_{am}(\mu) t_{bc}^{mk}
+ \tfrac{1}{2} \bar{h}_{ca}^{ke} r_{eb}^{}(\mu)
- \tfrac{1}{6} \bar{h}_{m}^{k} r_{abc}^{\phantom{ij}m}(\mu)
+ \tfrac{1}{2} \bar{h}_{c}^{e} r_{abe}^{\phantom{ij}k}(\mu)
+ \tfrac{1}{4} \bar{h}_{ab}^{ef} r_{efc}^{\phantom{ij}k}(\mu)
\nonumber \\
&
+ \tfrac{1}{2} \bar{h}_{cm}^{ke} r_{abe}^{\phantom{ij}m}(\mu)
+ \tfrac{1}{6} \bar{h}_{m}^{e} r_{abce}^{\phantom{ij}km}(\mu)
+ \tfrac{1}{4} \bar{h}_{cn}^{ef}r_{abef}^{\phantom{ij}kn}(\mu)
- \tfrac{1}{12} \bar{h}_{mn}^{kf} r_{abcf}^{\phantom{ij}mn}(\mu)
\nonumber \\
&
+ \tfrac{1}{12} I_{mn}(\mu)t_{abc}^{mnk}
],
\label{eq3p1h}
\end{align}
and
\begin{align}
\langle \Phi_{\phantom{ij}kl}^{abcd} | (\overline{H}_{N,\mathrm{open}}^{(\text{CCSDT})} R_{\mu}^{(+2)})_C | \Phi\rangle
= &\mathscr{A}_{abcd}\mathscr{A}^{kl}[
- \tfrac{1}{12}\bar{h}_{dm}^{lk} r_{abc}^{\phantom{kl}m}(\mu)
+ \tfrac{1}{4} \bar{h}_{dc}^{le} r_{abe}^{\phantom{ij}k}(\mu)
+ \tfrac{1}{12}I_{abc}^{\phantom{kl}e}(\mu) t_{ed}^{kl}
- \tfrac{1}{4} I_{abm}^{\phantom{ij}k}(\mu) t_{cd}^{ml}
- \tfrac{1}{24} \bar{h}_{m}^{l} r_{abcd}^{\phantom{ij}km}(\mu)
\nonumber \\
&
+ \tfrac{1}{12} \bar{h}_{a}^{e} r_{ebcd}^{\phantom{ij}kl}(\mu)
+ \tfrac{1}{16} \bar{h}_{ab}^{ef} r_{efcd}^{\phantom{ij}kl}(\mu)
+ \tfrac{1}{96} \bar{h}_{mn}^{kl} r_{abcd}^{\phantom{ij}mn}(\mu)
+ \tfrac{1}{6} \bar{h}_{dm}^{le} r_{abce}^{\phantom{ij}km}(\mu)
\nonumber \\
&
- \tfrac{1}{12} I_{am}(\mu) t_{bcd}^{mkl}
+ \tfrac{1}{8} I_{abe}^{\phantom{ij}m}(\mu) t_{cde}^{klm}
+ \tfrac{1}{12} I_{mnc}^{\phantom{ij}k}(\mu) t_{abd}^{mnl}
].
\label{eq4p2h}
\end{align}
The analogous expressions used to construct the
full DIP-EOMCCSDT(4$h$-2$p$)
eigenvalue problem associated with projections onto 
2$h$ ($|\Phi_{ij}\rangle$), 3$h$-1$p$ ($| \Phi_{ijk}^{\phantom{ab}c} \rangle$), and 4$h$-2$p$ ($|\Phi_{ijkl}^{\phantom{ab}cd} \rangle$) determinants are given by
(cf. Ref. \cite{DIP-EOMCCSDT})
\begin{equation}
\langle \Phi_{ij} | (\overline{H}_{N,\mathrm{open}}^{(\text{CCSDT})} R_{\mu}^{(-2)})_C | \Phi\rangle
= \mathscr{A}^{ij}[
-\bar{h}_{m}^{i} r_{}^{mj}(\mu)
+ \tfrac{1}{4} \bar{h}_{mn}^{ij}r_{}^{mn}(\mu)
+ \tfrac{1}{2} \bar{h}_{m}^{e}r_{\phantom{ab}e}^{ijm}(\mu)
- \tfrac{1}{2} \bar{h}_{mn}^{if}r_{\phantom{ab}f}^{mjn}(\mu)
+ \tfrac{1}{8} \bar{h}_{mn}^{ef}r_{\phantom{ab}ef}^{ijmn}(\mu)],
\label{eq2h}
\end{equation}
\begin{align}
\langle \Phi_{ijk}^{\phantom{ab}c} | (\overline{H}_{N,\mathrm{open}}^{(\text{CCSDT})} R_{\mu}^{(-2)})_C | \Phi\rangle
=&\mathscr{A}^{ijk}[
\tfrac{1}{2} {I^{\prime}}^{ie}(\mu) t_{ec}^{jk}
- \tfrac{1}{2} \bar{h}_{cm}^{ki} r_{}^{mj}(\mu)
+ \tfrac{1}{6} \bar{h}_{c}^{e} r_{\phantom{ab}e}^{ijk}(\mu)
- \tfrac{1}{2} \bar{h}_{m}^{k} r_{\phantom{ab}c}^{ijm}(\mu)
+ \tfrac{1}{4} \bar{h}_{mn}^{ij} r_{\phantom{ab}c}^{mnk}(\mu)
\nonumber \\
&+ \tfrac{1}{2} \bar{h}_{cm}^{ke} r_{\phantom{ab}e}^{ijm}(\mu)
+ \tfrac{1}{6} \bar{h}_{m}^{e} r_{\phantom{ab}ce}^{ijkm}(\mu)
- \tfrac{1}{4} \bar{h}_{mn}^{kf} r_{\phantom{ab}cf}^{ijmn}(\mu)
+ \tfrac{1}{12} \bar{h}_{cn}^{ef} r_{\phantom{ab}ef}^{ijkn}(\mu)
\nonumber \\
&+ \tfrac{1}{12} I^{ef}(\mu) t_{efc}^{ijk}],
\label{eq3h1p}
\end{align}
and
\begin{align}
\langle \Phi_{ijkl}^{\phantom{ab}cd} | (\overline{H}_{N,\mathrm{open}}^{(\text{CCSDT})} R_{\mu}^{(-2)})_C | \Phi\rangle
= &\mathscr{A}^{ijkl}\mathscr{A}_{cd}[
\tfrac{1}{12} \bar{h}_{dc}^{le} r_{\phantom{ab}e}^{ijk}(\mu)
- \tfrac{1}{4} \bar{h}_{dm}^{lk} r_{\phantom{ab}c}^{ijm}(\mu)
- \tfrac{1}{12} I_{\phantom{ab}m}^{ijk}(\mu) t_{cd}^{ml}
+ \tfrac{1}{4} I_{\phantom{ab}c}^{ije}(\mu) t_{ed}^{kl}
+ \tfrac{1}{24} \bar{h}_{d}^{e} r_{\phantom{ab}ce}^{ijkl}(\mu)
\nonumber \\
&- \tfrac{1}{12} \bar{h}_{m}^{i} r_{\phantom{ab}cd}^{mjkl}(\mu)
+ \tfrac{1}{16} \bar{h}_{mn}^{ij} r_{\phantom{ab}cd}^{mnkl}(\mu)
+ \tfrac{1}{96} \bar{h}_{cd}^{ef} r_{\phantom{ab}ef}^{ijkl}(\mu)
+ \tfrac{1}{6} \bar{h}_{dm}^{le} r_{\phantom{ab}ce}^{ijkm}(\mu)
\nonumber \\
&+ \tfrac{1}{12} I^{ie}(\mu) t_{ecd}^{jkl}
+ \tfrac{1}{8} I_{\phantom{ab}e}^{ijm}(\mu) t_{cde}^{klm}
+ \tfrac{1}{12} I_{\phantom{ab}c}^{efk}(\mu) t_{efd}^{ijl}].
\label{eq4h2p}
\end{align}
In writing Eqs.\ (\ref{eq2p})--(\ref{eq4h2p}), 
we have employed the Einstein summation convention over repeated upper and lower indices and made use of the antisymmetrization operators $\mathscr{A}^{pq} = \mathscr{A}_{pq} = 1 - (pq)$, $\mathscr{A}^{pqr} = \mathscr{A}^{p/qr}\mathscr{A}^{qr}$, and $\mathscr{A}^{pqrs} = \mathscr{A}^{p/qrs}\mathscr{A}^{qrs}$, with $\mathscr{A}^{p/qr} = 1 - (pq) - (pr)$ and $\mathscr{A}^{p/qrs} = 1 - (pq) - (pr) - (ps)$. Programmable expressions for the one-body ($\bar{h}_{p}^{q}$) and two-body ($\bar{h}_{pq}^{rs}$)
components of $\overline{H}_{N,\mathrm{open}}^{(\text{CCSDT})} = (H_{N} e^{T_1 + T_2 + T_3})_{C,\text{open}}$ 
can be found in Ref. \cite{DIP-EOMCCSDT}, while the additional intermediate quantities entering Eqs. (\ref{eq3p1h}), (\ref{eq4p2h}), (\ref{eq3h1p}), and (\ref{eq4h2p}) are provided in Table \ref{i-deadip}.

When implemented using the above expressions, the diagonalization 
steps characterizing DEA-EOMCCSDT(4$p$-2$h$) 
[Eqs.\ (\ref{eq2p})--(\ref{eq4p2h})] and 
DIP-EOMCCSDT(4$h$-2$p$) [Eqs.\ (\ref{eq2h})--(\ref{eq4h2p})] involve
computational steps that scale
as $n_o^2 n_u^6$ and $n_o^4 n_u^4$, respectively, where $n_o$ ($n_u$) denotes the number of occupied (unoccupied) spin-orbitals 
in $|\Phi\rangle$. 
Although the diagonalization costs of DEA-EOMCCSDT(4$p$-2$h$)/DIP-EOMCCSDT(4$h$-2$p$) are identical to those
of the lower-level DEA-EOMCCSD(4$p$-2$h$)/DIP-EOMCCSD(4$h$-2$p$) methods, the overall costs are higher 
because the underlying $N$-electron reference calculation involves solving the
CCSDT rather than CCSD equations, increasing the scaling of the ground-state
step from $n_o^2 n_u^4$ to $n_o^3 n_u^5$.
For DIP-EOMCCSDT(4$h$-2$p$), the CCSDT calculation is more expensive than the
$n_o^4 n_u^4$ diagonalization step and thus constitutes the dominant 
computational cost, rendering DIP-EOMCCSDT(4$h$-2$p$) substantially more
demanding than DIP-EOMCCSD(4$h$-2$p$) in practice. 
For DEA-EOMCCSDT(4$p$-2$h$), the diagonalization step remains the
dominant cost, just as in DEA-EOMCCSD(4$p$-2$h$),
but this does not
lessen the severe practical impact of the
$n_{u}^6$-scaling
diagonalization within the 4$p$-2$h$ sector. 
In realistic applications employing larger basis sets, where $n_u \gg n_o$, 
full DEA-EOMCCSDT(4$p$-2$h$) calculations are feasible only for 
very small systems.

As mentioned in the Introduction, our main objective is to use 
active-space CC ideas to construct robust approximations to the prohibitively
expensive DEA-EOMCCSDT(4$p$-2$h$) and DIP-EOMCCSDT(4$h$-2$p$) methods. In the 
active-space DEA/DIP-EOMCC approaches developed in this work, we follow the strategy 
previously employed in the CCSDt \cite{semi0b,semi2,ccsdtq3,semi4,piecuch-qtp} and
DEA-EOMCCSD(4$p$-2$h$)$\{N_{u}\}$/DIP-EOMCCSD(4$h$-2$p$)$\{N_{o}\}$ \cite{jspp-dea-dip-2013, jspp-dea-dip-2014, jspp-dea-dip-2017, jspp-dea-dip-2021}
formalisms. Thus, we constrain the excitations entering the $T_{3}$, 
$R_{\mu,4p\mbox{-}2h}$, and $R_{\mu,4h\mbox{-}2p}$ operators  using active orbitals as follows:
\begin{equation}
t_{3} =
\sum_{
i < j < {\bf K}, {\bf A} < b < c
}
t_{{\bf A}bc}^{ij{\bf K}}
\; a^{\bf A} a^{b} a^{c} a_{\bf K} a_{j} a_{i},
\label{t3active}
\end{equation}
\begin{equation}
r_{\mu,4p\mbox{-}2h} =
\sum_{
k < l, {\bf A} < {\bf B} < c < d
}
r_{{\bf AB}cd}^{\;\;\;\;\;\; kl}(\mu) 
\; a^{\bf A} a^{\bf B} a^{c} a^{d} a_{l} a_{k},
\label{4p2hactive}
\end{equation}
and
\begin{equation}
r_{\mu,4h\mbox{-}2p} = \sum_{
i < j < {\bf K} < {\bf L}, c < d
}
r^{ij{\bf KL}}_{\;\;\; cd}(\mu) 
\; a^{c} a^{d} a_{\bf L} a_{\bf K} a_{j} a_{i}.
\label{4h2pactive}
\end{equation}
Here, bold uppercase indices ${\bf A}$ and ${\bf B}$ denote active 
spin-orbitals unoccupied in the $N$-electron reference $|\Phi\rangle$, 
whereas ${\bf K}$ and ${\bf L}$ denote active spin-orbitals occupied in 
$|\Phi\rangle$. When the truncated forms of the doubly electron attaching and 
doubly ionizing operators defined by Eqs.\ (\ref{4p2hactive}) and 
(\ref{4h2pactive}) are used in combination with the full $T_{3}$ operator, 
one obtains the CCSDT-based DEA-EOMCCSDT(4$p$-2$h$)$\{N_{u}\}$ and 
DIP-EOMCCSDT(4$h$-2$p$)$\{N_{o}\}$ methods,
respectively. 
When we replace $T_{3}$
with $t_{3}$, i.e.,
we use all three
Eqs.\ (\ref{t3active})--(\ref{4h2pactive}),
we obtain their more economical CCSDt-based DEA-EOMCCSDt(4$p$-2$h$)$\{N_{u}\}$ 
and DIP-EOMCCSDt(4$h$-2$p$)$\{N_{o}\}$ counterparts. Regardless of whether the 
full $T_{3}$ operator or its active-space $t_{3}$ counterpart is employed, 
the active-space restrictions on the 4$p$-2$h$/4$h$-2$p$ excitations reduce 
the costs of the diagonalization steps to the much more affordable 
$N_{u}^{2} n_{o}^{2} 
n_{u}^{4}$ 
for 
DEA-EOMCCSDt(4$p$-2$h$)\{$N_{u}$\}/DEA-EOMCCSDT(4$p$-2$h$)\{$N_{u}$\}
and
$N_{o}^{2} n_{o}^{2} n_{u}^{4}$ for DIP-EOMCCSDt(4$h$-2$p$)\{$N_{o}$\}/DIP-EOMCCSDT(4$h$-2$p$)\{$N_{o}$\}. In addition, for DEA-EOMCCSDt(4$p$-2$h$)\{$N_{u}$\} and
DIP-EOMCCSDt(4$h$-2$p$)\{$N_{o}$\}, the preceding $N$-electron CCSDt step 
can be performed using $N_{o} N_{u} n_{o}^{2} n_{u}^{4}$ operations, 
which amount
to a small prefactor
times the costs of CCSD.
\section{Results and Discussion}
\label{sec3}
%
%
The calculations presented in this section are designed to assess the
performance of the newly developed DEA/DIP-EOMCC approaches that
incorporate full and active-space treatments of 4$p$-2$h$/4$h$-2$p$
excitations and three-body clusters. Building on our earlier work
\cite{jspp-dea-dip-2013, jspp-dea-dip-2014, jspp-dea-dip-2017, jspp-dea-dip-2021},
which demonstrated that DEA- and DIP-EOMCCSD methods with full and
active-space 4$p$-2$h$ and 4$h$-2$p$ treatments provide a highly accurate
description of MR systems containing two electrons or two holes outside
a closed-shell core, substantially outperforming their counterparts
truncated at the 3$p$-1$h$ and 3$h$-1$p$ levels, the present study
addresses two key questions: (i) does the inclusion of
$T_{3}$ or $t_{3}$ clusters improve the results obtained with 
the DEA-EOMCCSD(4$p$-2$h$) and DIP-EOMCCSD(4$h$-2$p$) methods and their active-space variants and (ii) do the active-space
DEA-EOMCCSDt(4$p$-2$h$)\{$N_{u}$\}/DEA-EOMCCSDT(4$p$-2$h$)\{$N_{u}$\}
and DIP-EOMCCSDt(4$h$-2$p$)\{$N_{o}$\}/DIP-EOMCCSDT(4$h$-2$p$)\{$N_{o}$\}
approaches recover
the energetics of their 
more expensive
DEA-EOMCCSDT(4$p$-2$h$) and DIP-EOMCCSDT(4$h$-2$p$) parents?

To answer these questions, we employ the highly efficient,
spin-integrated
DEA-EOMCCSDt(4$p$-2$h$)\{$N_{u}$\},
DEA-EOMCCSDT(4$p$-2$h$)\{$N_{u}$\},
DEA-EOMCCSDT(4$p$-2$h$),
DIP-EOMCCSDt(4$h$-2$p$)\{$N_{o}$\},
DIP-EOMCCSDT(4$h$-2$p$)\{$N_{o}$\}, and
DIP-EOMCCSDT(4$h$-2$p$) 
codes developed in this study 
within the GAMESS package
\cite{gamess2020,gamess2023}
using the same
automated formula derivation and implementation software as that exploited in our previous full and active-space DEA-EOMCCSD(4$p$-2$h$)/DIP-EOMCCSD(4$h$-2$p$) work 
\cite{jspp-dea-dip-2013, jspp-dea-dip-2014, jspp-dea-dip-2017, jspp-dea-dip-2021}.
As with the existing suite of DEA/DIP-EOMCC methods
that we have incorporated in GAMESS in recent years, the new active-space
DEA-EOMCCSDt(4$p$-2$h$)\{$N_{u}$\}/DIP-EOMCCSDt(4$h$-2$p$)\{$N_{o}$\}
and DEA-EOMCCSDT(4$p$-2$h$)\{$N_{u}$\}/DIP-EOMCCSDT(4$h$-2$p$)\{$N_{o}$\}
approaches, as well as their
DEA-EOMCCSDT(4$p$-2$h$)/DIP-EOMCCSDT(4$h$-2$p$) parents, were
interfaced with the integral, restricted
\linebreak
Hartree--Fock (RHF), restricted open-shell Hartree--Fock (ROHF), integral transformation, and CC (CCSD, CCSDt, and CCSDT) 
GAMESS modules. Using these implementations, we computed the adiabatic energy
gaps between the triplet ground state ($X \; ^{3}B_{1}$) and the three
lowest-energy singlet excited states ($A \; ^{1}A_{1}$,
$B \; ^{1}B_{1}$, and $C \; ^{1}A_{1}$) of methylene described by the TZ2P basis set
of Ref.\ \cite{ch2tz2p} (obtained by augmenting the triple-zeta basis of Ref.\ \cite{tz2p} with two sets of polarization functions) and the adiabatic singlet--triplet separation
in TMM described with the cc-pVDZ basis \cite{ccpvnz}, comparing them with the available
full CI \cite{ch2tz2p} and experimentally derived \cite{TMM-Expt.,ch2_krylov}
data. We also compare the vertical DIPs corresponding
to the lowest singlet and
triplet dicationic states of the 23 atoms and molecules
studied
in Ref. \cite{loos-ppbse} using
CIPSI calculations
extrapolated to the 
full CI limit
and the aug-cc-pVTZ basis
\cite{ccpvnz,augccpvnz} with their counterparts
obtained using the various
DIP-EOMCC methods discussed in this Letter.
While our codes allow various orbital choices, including, for example, the HF orbitals optimized for the
$(N \pm 2)$-electron target systems, all
DEA/DIP-EOMCC calculations reported in
this work, along with the underlying CC computations, were performed
using the RHF orbitals obtained for the corresponding $N$-electron
closed-shell reference species.
One might argue that such a choice of molecular orbitals may introduce numerical problems in
the DIP-EOMCC computations when the underlying dianionic $N$-electron reference species becomes
unstable with respect to electron detachment, prompting the use of the recipes considered in
Refs.\ \cite{kus-krylov-2011,kus-krylov-2012}, but this is not an issue in this work, where we do
not use basis sets with highly diffuse functions. Furthermore, as shown in Ref.\ \cite{jspp-dea-dip-2014},
problems of this type largely disappear once the high-order 4$h$-2$p$ excitations are explicitly
included in the calculations, as is the case in the present study, and we always have the option of
turning to the RHF or ROHF orbitals of the target $(N - 2)$-electron system. It is worth emphasizing
in this context that the DEA-EOMCC approaches do not face such issues. It is, therefore, useful to
develop them alongside their DIP-EOMCC counterparts, which is yet another reason why both of
these methodologies are pursued in this work.
With the exception of the DIP-EOMCC
calculations of LiF and BeO, the lowest-energy orbitals corresponding to
chemical cores of non-hydrogen atoms were kept frozen in the post-HF steps
of the DEA/DIP-EOMCC computations. 
In the case of LiF
and BeO, we followed Ref.\ \cite{loos-ppbse} and correlated the $1s$ orbitals of Li and Be.

%
We begin by examining the DEA- and DIP-EOMCC adiabatic energy gaps corresponding to the low-lying states of methylene. As shown in Table \ref{CH2}, both the full and active-space DEA-EOMCCSD(4$p$-2$h$) and DIP-EOMCCSD(4$h$-2$p$) approaches reproduce the exact, full CI results for the
$A \; ^{1}A_{1} - X \; ^{3}B_{1}$,
$B \; ^{1}B_{1} - X \; ^{3}B_{1}$,
and 
$C \; ^{1}A_{1} - X \; ^{3}B_{1}$
singlet--triplet separations with the small mean unsigned error (MUE) and nonparallelity error (NPE) values
on the order of a few tenths of a kcal/mol representing
major improvements over the 
DEA-EOMCCSD(3$p$-1$h$)/DIP-EOMCCSD(3$h$-1$p$) data.
While the incorporation of 4$p$-2$h$/4$h$-2$p$ correlations
is essential for obtaining quantitative results, inclusion of triple excitations in the cluster operator through the active-space CCSDt method or full CCSDT does not seem to have
a substantial impact on the
full or active-space
DEA-EOMCCSD(4$p$-2$h$)/DIP-EOMCCSD(4$h$-2$p$) data.
The insignificant effect of the three-body clusters on the singlet--triplet gaps
shown in Table \ref{CH2}
can be understood if we
realize that we do not
change the number of electrons when examining different electronic states of a
given species, so that the states being compared, which belong to the same ($N+2$)-electron (DEA-EOMCC) or ($N-2$)-electron (DIP-EOMCC) sector of the Fock space, are treated on equal footing and the $T_{3}$ 
contributions may largely cancel out.
Most importantly, the relatively inexpensive active-space formulations of all DEA-EOMCC and DIP-EOMCC approaches
considered in this work
employing small numbers of active orbitals
closely reproduce the corresponding parent data for all states of methylene 
shown in Table \ref{CH2}. 
In particular, the CCSD-based DEA-EOMCCSD(4$p$-2$h$)$\{2\}$ and DIP-EOMCCSD(4$h$-2$p$)$\{2\}$ approaches, as well as their only moderately more expensive CCSDt-based DEA-EOMCCSDt(4$p$-2$h$)$\{2\}$ and DIP-EOMCCSDt(4$h$-2$p$)$\{2\}$ counterparts, provide similar accuracies to the orders-of-magnitude more expensive CCSDT-based DEA-EOMCCSDT(4$p$-2$h$) and DIP-EOMCCSDT(4$h$-2$p$) calculations.

%
We next consider the DEA-EOMCC and DIP-EOMCC calculations of the adiabatic singlet--triplet separation in TMM using the cc-pVDZ basis, summarized in Table \ref{TMM}. It is well established that the TMM's ground state at its equilibrium $D_{3h}$-symmetric geometry, abbreviated as $X\,^{3}A_{2}^{\prime}$, is a triplet (see, e.g., Refs.\ \cite{ch2_krylov,tmm_krylov,jspp-dea-dip-2013}). At this geometry, there also exists a low-lying pair of degenerate singlet states of $E^{\prime}$ symmetry, which, after being stabilized by the Jahn--Teller distortion that lifts their degeneracy, correlate with the 
$C_{s}$-symmetric
open-shell singlet that
can be approximated by a twisted
$C_{2v}$ structure designated by
$A\,^{1}B_{1}$ and the 
$C_{2v}$-symmetric
multiconfigurational 
$B\,^{1}A_{1}$ state that
has been detected via photoelectron spectroscopy in Ref.\ \cite{TMM-Expt.}.
Accordingly, in Table \ref{TMM} we report the adiabatic singlet--triplet gaps, $\Delta E_{\rm S-T}=E_S-E_T$, obtained with the various full and active-space DEA/DIP-EOMCC approaches, where $E_S$ is the energy of the $B\,^{1}A_{1}$ state in its $C_{2v}$-symmetric structure
and $E_T$ is the energy of the $X\,^{3}A_{2}^{\prime}$ 
ground state in its $D_{3h}$-symmetric minimum. The geometries characterizing
the $B\,^{1}A_{1}$ and $X\,^{3}A_{2}^{\prime}$ states
of TMM used in this study were taken from 
Ref.\ \cite{ch2_krylov}.
In analogy to methylene, the results reported in
Table \ref{TMM} show that the
inclusion of 4$p$-2$h$/4$h$-2$p$
correlations in the DEA/DIP-EOMCC
calculations is essential for obtaining a
highly accurate description of the
singlet--triplet separation in TMM, but $T_{3}$ contributions to the cluster operator can be ignored. Indeed, 
the singlet--triplet gaps calculated using the full and active-space DEA-EOMCCSD(4$p$-2$h$)/DIP-EOMCCSD(4$h$-2$p$) methods yield $\Delta E_{\rm S-T}$ values within  $\sim$1 kcal/mol of the experimentally derived purely electronic singlet--triplet gap (obtained by removing the zero-point vibrational contributions estimated in Ref.\ \cite{ch2_krylov} from the experimental $\Delta E_{\rm S-T}$ reported in Ref.\ \cite{TMM-Expt.}), whereas the 
DEA-EOMCCSD(3$p$-1$h$)/DIP-EOMCCSD(3$h$-1$p$)
gaps are $\sim$2--5 kcal/mol higher. Replacing the CCSD description of the underlying $N$-electron $({\rm TMM})^{2+}$/$({\rm TMM})^{2-}$ species in the full or active-space DEA-EOMCCSD(4$p$-2$h$)/DIP-EOMCCSD(4$h$-2$p$) calculations with CCSDt or CCSDT changes the resulting $\Delta E_{\rm S-T}$ values by $\sim$0.2 kcal/mol or less, indicating that the 
contributions to the energies of the $X\,^{3}A_{2}^{\prime}$ and
$B\,^{1}A_{1}$ states due to the $T_{3}$ clusters 
largely cancel out when considering the $B\,^{1}A_{1} - X\,^{3}A_{2}^{\prime}$ separation.
It is important to note though that the active-space DEA/DIP-EOMCC
results accurately approximate 
their full-space
parents, even when the numbers of active orbitals used in the calculations are small. In particular, the
active-space treatment of
4$p$-2$h$/4$h$-2$p$ excitations on top of CCSD, CCSDt, or CCSDT produces
$\Delta E_{\rm S-T}$ values in the
19.00--19.32 kcal/mol range, in excellent agreement
with 19.19--19.68 kcal/mol obtained
using a much more expensive 
full treatment of 
4$p$-2$h$/4$h$-2$p$ correlations. 
In analogy to methylene, the DEA/DIP-EOMCC 
calculations based on small numbers of active orbitals to
identify the dominant 4$p$-2$h$/4$h$-2$p$
and, optionally, $T_{3}$ contributions provide
practical
routes for a highly accurate determination of the 
adiabatic singlet--triplet 
gap in TMM.

%
A somewhat different pattern emerges for the vertical DIPs of the 23 atoms and molecules summarized in Table \ref{23}, where the DIP-EOMCC results obtained using the aug-cc-pVTZ basis set are compared with their near-exact counterparts reported in Ref. \cite{loos-ppbse} based on extrapolating CIPSI energetics to the full CI limit. While extending the correlation treatment from DIP-EOMCCSD(3$h$-1$p$) to DIP-EOMCCSD(4$h$-2$p$) leads to significant improvements, lowering the MUEs relative to the extrapolated CIPSI DIP values corresponding to the lowest singlet and triplet dicationic states from $\sim$0.6 eV to $\sim$0.2 eV, the inclusion of $T_{3}$ clusters in the ground-state CC calculations for the underlying $N$-electron species produces further substantial error reductions, which are not seen in the calculations of the singlet--triplet gaps in methylene and TMM.
Indeed, the perturbative [CCSD(T)(a)], active-space (CCSDt), and full (CCSDT) treatments of the triply excited clusters reduce the MUEs characterizing the DIP-EOMCCSD(4$h$-2$p$) calculations by factors of about 2--5. This is because, in contrast to the previously examined singlet--triplet splittings, the vertical DIPs considered in Table \ref{23} represent energy differences involving the $N$- and $(N-2)$-electron states. As a result, the $T_{3}$ contributions to the neutral and doubly ionized states do not cancel out and, consistent with the findings reported in Ref.\ \cite{DIP-EOMCCSDT}, become important for obtaining a balanced treatment of the $N$- and $(N-2)$-electron species. In this context, it is especially encouraging that the active-space DIP-EOMCCSDt(4$h$-2$p$)\{$N_o$\} approach is as accurate as its much more expensive full DIP-EOMCCSDT(4$h$-2$p$) parent, with the corresponding 
mean signed error (MSE), MUE, and maximum unsigned error (MaxE) values
relative to the CIPSI-based benchmarks not exceeding 0.1 eV.
Furthermore, while the overall
performance of the active-space
DIP-EOMCCSDt(4$h$-2$p$)\{$N_o$\} approach, measured by the MSEs and MUEs, is
comparable to that of the perturbative DIP-EOMCCSD(T)(a)(4$h$-2$p$) method
introduced in Ref.\ \cite{DIP-EOMCCSDT}, the former approach reduces the MaxE values
obtained with the latter method from 0.43 eV to 0.17--0.18 eV (for a complete set of the DIPs corresponding to the lowest singlet and triplet dicationic states of the 23 atoms and
molecules studied in Ref.\ \cite{loos-ppbse} obtained in the DIP-EOMCC computations performed in this work and the information about the numbers of active orbitals used in the DIP-EOMCCSDt(4$h$-2$p$)\{$N_o$\} calculations, see Tables S1--S3 of the Supplementary Data). Our calculations clearly indicate that the active-space DIP-EOMCCSDt(4$h$-2$p$)\{$N_{o}$\} methodology provides a promising, computationally practical tool for high-accuracy calculations of DIPs.

\section{Summary}
\label{sec4}
We have efficiently implemented and tested the full and active-space variants of the DEA-EOMCCSDT(4$p$-2$h$) and DIP-EOMCCSDT(4$h$-2$p$) approaches, assessing their performance using the low-lying states of methylene, the singlet--triplet gap of TMM, and the vertical DIPs of 23 atoms and molecules as representative examples. Our computations demonstrate that 4$p$-2$h$/4$h$-2$p$ correlations are essential for obtaining a quantitative description in all studied cases, but $T_{3}$ effects appear to be less significant unless energy differences involving species with different numbers of electrons, such as DIPs, are considered.
Across all examined systems, the DEA/DIP-EOMCC formulations with an active-space treatment of 4$p$-2$h$/4$h$-2$p$ excitations on top of CCSD, CCSDt, and CCSDT, using small numbers of active orbitals to identify the leading 4$p$-2$h$/4$h$-2$p$ and, in the case of CCSDt, $T_{3}$ amplitudes, closely reproduce their parent full-space counterparts while offering substantial savings in the computational effort.
Thus, although full DEA-EOMCCSDT(4$p$-2$h$) and DIP-EOMCCSDT(4$h$-2$p$) implementations are now available \cite{DIP-EOMCCSDT,deprince-dipeomccsdt,krylov-dipea-eomccsdt}, the present results demonstrate that the corresponding active-space methods provide a much more practical route for routine applications of the high-level approaches in these categories, retaining their accuracy at small fractions of the computational costs.
\section*{CRediT authorship contribution statement}
\textbf{Jun Shen}: Methodology, Software, Data curation, Formal
analysis, Investigation, Validation, Writing - original draft.
\textbf{Karthik Gururangan}: Methodology, Software, Data curation, Formal analysis, Investigation, Validation, Writing - original draft.
\textbf{Piotr Piecuch}: Conceptualization, Methodology, Formal
analysis, Investigation, Funding acquisition, Project administration,
Resources, Supervision, Validation, Writing - reviewing and editing.
\section*{Declaration of competing interest}
The authors declare that they have no known competing financial
interests or personal relationships that could have appeared to influence
the work reported in this paper.
\section*{Data availability}
The data that support the findings of this study are available
within the article and the Supplementary Data.
\section*{Acknowledgments}
This work has been supported by the Chemical Sciences,
Geosciences and Biosciences Division,
Office of Basic Energy Sciences,
Office of Science,
U.S. Department of Energy
(Grant No. DE-FG02-01ER15228 to P.P.).
\section*{Appendix A. Supplementary data}
Supplementary material associated with this article can be found online at
%
\bibliographystyle{elsarticle-num}
\biboptions{sort,compress}
\bibliography{refs-v2}

\newpage
\clearpage
\pagebreak
\begin{table*}[h]
\begin{threeparttable}
\caption{
The intermediates entering Eqs. (\ref{eq3p1h}), (\ref{eq4p2h}), (\ref{eq3h1p}), and (\ref{eq4h2p})
that are introduced in order to evaluate the contributions to the DEA-EOMCCSDT(4$p\mbox{-}2h$) and DIP-EOMCCSDT(4$h\mbox{-}2p$) equations due to the three- and four-body components of the similarity-transformed Hamiltonian.
}
\label{i-deadip}
\begin{tabular}{cc}
\hline\hline
Intermediate & Expression\tnote{a}\\
\hline
DEA-EOMCCSDT(4$p\mbox{-}2h$)\\
${I^{\prime}}_{am}(\mu)$ & $\tfrac{1}{2}\bar{h}_{am}^{ef} r_{ef}^{}(\mu) + \tfrac{1}{2}\bar{h}_{nm}^{ef} r_{afe}^{\phantom{ij}n}(\mu)$ \vspace{0.3em}\\
$I_{am}(\mu)$& $\tfrac{1}{2}\bar{h}_{am}^{ef} r_{ef}^{}(\mu) + \tfrac{1}{2}\bar{h}_{nm}^{ef} r_{afe}^{\phantom{ij}n}(\mu) + \bar{h}_{m}^{e} r_{ae}^{}(\mu)$ \vspace{0.3em}\\
$I_{mn}(\mu)$ & $\tfrac{1}{2} \bar{h}_{mn}^{ef} r_{ef}^{}(\mu)$ \vspace{0.3em}\\
$I_{abc}^{\phantom{ij}e}(\mu)$ & $\mathscr{A}_{abc}[\tfrac{1}{2} \bar{h}_{cm}^{fe} r_{abf}^{\phantom{ij}m}(\mu) - \tfrac{1}{2} \bar{h}_{ac}^{fe} r_{fb}^{}(\mu) - \tfrac{1}{12} \bar{h}_{mn}^{ef} r^{\phantom{ij}mn}_{abcf}(\mu)]$ \vspace{0.3em}\\
$I_{abm}^{\phantom{ij}k}(\mu)$ & $ \bar{h}_{nm}^{ke} r_{abe}^{\phantom{ij}m}(\mu) - \tfrac{1}{2} I_{nm}(\mu) t_{ab}^{kn} - \mathscr{A}_{ab}[\bar{h}_{am}^{ke} r_{eb}^{}(\mu) - \tfrac{1}{2} \bar{h}_{am}^{fe} r_{fbe}^{\phantom{ij}k}(\mu)] + \tfrac{1}{2} \bar{h}_{mn}^{ef} r_{abef}^{\phantom{ij}kn}(\mu)$ \vspace{0.3em}\\
$I_{abe}^{\phantom{ij}m}(\mu)$& $\bar{h}_{mn}^{ef} r_{abf}^{\phantom{ij}n}(\mu) + \mathscr{A}_{ab} \bar{h}^{fe}_{bm} r_{af}^{}(\mu)$ \vspace{0.3em}\\
$I_{mnc}^{\phantom{ij}k}(\mu)$& $\tfrac{1}{2} \bar{h}_{mn}^{ef} r_{efc}^{\phantom{ij}k}(\mu) - \bar{h}_{mn}^{ke} r_{ce}^{}(\mu)$ \vspace{0.3em}\\
DIP-EOMCCSDT(4$h\mbox{-}2p$)\tnote{b}\\
${I^{\prime}}^{ie}(\mu)$ & $\tfrac{1}{2}\bar{h}_{mn}^{ie} r^{mn}(\mu) - \tfrac{1}{2}\bar{h}_{mn}^{fe} r_{\phantom{ab}f}^{inm}(\mu)$ \vspace{0.3em}\\
$I^{ie}(\mu)$& $\tfrac{1}{2}\bar{h}_{mn}^{ie} r^{mn}(\mu) - \tfrac{1}{2}\bar{h}_{mn}^{fe} r_{\phantom{ab}f}^{inm}(\mu) - \bar{h}_{m}^{e} r^{im}(\mu)$ \vspace{0.3em}\\
$I^{ef}(\mu)$ & $\tfrac{1}{2} \bar{h}_{mn}^{ef} r^{mn}(\mu)$ \vspace{0.3em}\\
$I_{\phantom{ab}m}^{ijk}(\mu)$ & $\mathscr{A}^{ijk}[\tfrac{1}{2} \bar{h}_{nm}^{ke} r_{\phantom{ab}e}^{ijn}(\mu) - \tfrac{1}{2} \bar{h}_{nm}^{ik} r^{nj}(\mu) + \tfrac{1}{12} \bar{h}_{mn}^{ef} r_{\phantom{ab}ef}^{ijkn}(\mu)]$ \vspace{0.3em}\\
$I_{\phantom{ab}c}^{ije}(\mu)$ & $\bar{h}_{cm}^{fe} r_{\phantom{ab}e}^{ijm}(\mu) + \tfrac{1}{2} I^{fe}(\mu) t_{cf}^{ij} + \mathscr{A}^{ij}[\bar{h}_{cm}^{ie} r^{mj}(\mu) + \tfrac{1}{2} \bar{h}_{nm}^{ie} r_{\phantom{ab}c}^{njm}(\mu)] - \tfrac{1}{2} \bar{h}_{mn}^{ef} r_{\phantom{ab}cf}^{ijmn}(\mu)$ \vspace{0.3em}\\
$I_{\phantom{ab}e}^{ijm}(\mu)$& $\bar{h}_{mn}^{ef} r_{\phantom{ab}f}^{ijn}(\mu) - \mathscr{A}^{ij} \bar{h}_{nm}^{je} r^{in}(\mu)$ \vspace{0.3em}\\
$I_{\phantom{ab}c}^{efk}(\mu)$& $\tfrac{1}{2} \bar{h}_{mn}^{ef} r_{\phantom{ab}c}^{mnk}(\mu) + \bar{h}_{cm}^{ef} r^{km}(\mu)$ \vspace{0.3em}\\
\hline\hline
\end{tabular}
\begin{tablenotes}
\footnotesize
\item[a]{
In each expression, summation is carried out over repeated upper and lower indices.
}
\item[b]{
From Ref. \cite{DIP-EOMCCSDT}.
}
\end{tablenotes}
\end{threeparttable}
\end{table*}

\newpage
\clearpage
\pagebreak
\begin{table*}[h]
\centering
\footnotesize
\begin{threeparttable}
\caption{
Comparison of the full CI and various DEA/DIP-EOMCC adiabatic excitation energies characterizing the low-lying states of methylene, as described using the TZ2P basis.
The basis set, geometries, and full CI values were taken from Ref. \cite{ch2tz2p}.
The full CI excitation energies are reported in kcal/mol, whereas all DEA/DIP-EOMCC results,
including the MUE and NPE values,
correspond to errors relative to full CI, also in kcal/mol. 
The DEA-EOMCC and DIP-EOMCC calculations employed the RHF orbitals of the underlying 
$({\rm CH}_{2})^{2+}$ (DEA) and $({\rm CH}_{2})^{2-}$ (DIP) reference species.
Following Ref. \cite{ch2tz2p},
the lowest-energy occupied and the highest-energy
unoccupied orbitals were frozen
in all post-RHF steps
and the spherical components of the carbon $d$ orbital were employed throughout.
}
\label{CH2}
\begin{tabular}{l @{\extracolsep{0.20in}} c @{\extracolsep{0.20in}} c @{\extracolsep{0.20in}} c
                  @{\extracolsep{0.20in}} c @{\extracolsep{0.25in}} c }
\hline\hline
Method & $A \; ^{1}A_{1} - X \; ^{3}B_{1}$  & $B \; ^{1}B_{1} - X \; ^{3}B_{1}$
& $C \; ^{1}A_{1} - X \; ^{3}B_{1}$ & MUE & NPE \\
\hline
DEA-EOMCCSD(3$p$-1$h$)                   &$-$0.11&$-$1.89&   $-$3.64&3.64&3.53\\
DEA-EOMCCSD(4$p$-2$h$)\{2\} \tnote{a}   &   0.13&$-$0.35&$-$0.54&0.54&0.67\\
DEA-EOMCCSD(4$p$-2$h$)                   &   0.38&$-$0.02&   0.22&0.38&0.40\\
DIP-EOMCCSD(3$h$-1$p$)                   &$-$4.53&$-$3.22&$-$4.63&4.63&1.41\\
DIP-EOMCCSD(4$h$-2$p$)\{2\} \tnote{b}   &$-$0.33&$-$0.49&$-$0.33&0.49&0.16\\
DIP-EOMCCSD(4$h$-2$p$)                   &$-$0.44&$-$0.51&$-$0.47&0.51&0.07\\[1mm]
DEA-EOMCCSDt(4$p$-2$h$)\{2\}\tnote{a,c}&$-$0.07&$-$0.39&$-$1.00&1.00&0.93\\
DIP-EOMCCSDt(4$h$-2$p$)\{2\}\tnote{b,d}&$-$0.32&$-$0.41&$-$0.33&0.41&0.09\\[1mm]
DEA-EOMCCSDT(4$p$-2$h$)\{2\}\tnote{a}   &$-$0.05&$-$0.40&$-$1.00&1.00&0.95\\
DEA-EOMCCSDT(4$p$-2$h$)                  &   0.20&$-$0.08&$-$0.26&0.26&0.46\\
DIP-EOMCCSDT(4$h$-2$p$)\{2\}\tnote{b}  &$-$0.48&$-$0.39&$-$0.36&0.48&0.12\\
DIP-EOMCCSDT(4$h$-2$p$)                  &$-$0.58&$-$0.40&$-$0.51&0.58&0.18\\
Full CI &  11.14 &  35.59 & 61.67\\
\hline\hline
\end{tabular}

\begin{tablenotes}
\footnotesize
\item[a]{
The active orbitals used to select the dominant 4$p$-2$h$
excitations in the active-space DEA-EOMCC calculations consisted of
the two lowest-energy unoccupied orbitals of the $({\rm CH}_{2})^{2+}$
reference system, $3a_{1}$ and $1b_{1}$, which correlate with the
highest-occupied ($3a_{1}$) and lowest-unoccupied ($1b_{1}$) molecular orbitals
of ${\rm CH}_{2}$.
}
\item[b]{
The active orbitals used to select the dominant 4$h$-2$p$
excitations in the active-space DIP-EOMCC calculations consisted of
the two highest-energy occupied orbitals of the $({\rm CH}_{2})^{2-}$
reference system, $3a_{1}$ and $1b_{1}$, which correlate with the
highest-occupied ($3a_{1}$) and lowest-unoccupied ($1b_{1}$) molecular orbitals
of ${\rm CH}_{2}$.
}
\item[c]{
The active orbitals used to define the ground-state CCSDt calculations
preceding the DEA-EOMCCSDt(4$p$-2$h$)$\{2\}$ diagonalizations consisted of the highest occupied
($1b_{2}$) and two lowest unoccupied ($3a_{1}$ and $1b_{1}$) orbitals of $({\rm CH}_{2})^{2+}$, which correlate with the two highest occupied ($1b_{2}$, $3a_{1}$) and the lowest unoccupied ($1b_{1}$) orbitals of ${\rm CH}_{2}$.
}
\item[d]{
The active orbitals used to define the ground-state CCSDt calculations
preceding the DIP-EOMCCSDt(4$h$-2$p$)$\{2\}$ diagonalizations consisted of the two
highest occupied ($3a_{1}$ and $1b_{1}$) and the lowest unoccupied ($4a_{1}$) orbitals of
$({\rm CH}_{2})^{2-}$, which correlate with the highest occupied ($3a_{1}$) and two lowest unoccupied ($1b_{1}$, $4a_{1}$) orbitals of ${\rm CH}_{2}$.
}
\end{tablenotes}
\end{threeparttable}
\end{table*}

\newpage
\clearpage
\pagebreak
\begin{table*}[h]
\centering
\footnotesize
\caption{
The various DEA/DIP-EOMCC results for the adiabatic singlet--triplet separation in TMM,
as described by the cc-pVDZ basis set,
\mbox{$\Delta E_{\rm S-T} = E(B \; {^{1}}A_{1}) - E(X \; {^{3}}A_{2}^{\prime})$},
in kcal/mol, obtained in this work using the SF-DFT/6-31G(d) geometries of the
$X \; {^{3}}A_{2}^{\prime}$ and $B \; {^{1}}A_{1}$ states reported in Ref.\ \cite{ch2_krylov}. 
The $({\rm TMM})^{2+}$ reference system employed in the DEA-EOMCC computations was obtained by
vacating the doubly degenerate valence $1e^{\prime\prime}$ (the $D_{3h}$-symmetric
$X \; {^{3}}A_{2}^{\prime}$ state) or non-degenerate $1a_{2}$ and $2b_{1}$ (the
$C_{2v}$-symmetric $B \; {^{1}}A_{1}$ state) orbitals of the TMM's $\pi$ network.
The $({\rm TMM})^{2-}$ reference system employed in the DIP-EOMCC computations was obtained by
filling the valence $1e^{\prime\prime}$ (the $D_{3h}$-symmetric
$X \; {^{3}}A_{2}^{\prime}$ state) or $1a_{2}$ and $2b_{1}$ (the
$C_{2v}$-symmetric $B \; {^{1}}A_{1}$ state) orbitals.
The lowest-energy core orbitals correlating with the $1s$ shells of the carbon atoms were frozen in the post-RHF calculations and the spherical components of the $d$ orbitals were used throughout.
}
\label{TMM}

\begin{threeparttable}
\renewcommand{\TPTminimum}{\textwidth}

\begin{minipage}{\textwidth}
\centering
\begin{tabular}{l @{\extracolsep{0.06in}} c}
\hline\hline
Method & $\Delta E_{\rm S-T}$\\
\hline
DEA-EOMCCSD(3$p$-1$h$)                    & 23.92\\
DEA-EOMCCSD(4$p$-2$h$)\{3\}\tnote{a}      & 19.00\\
DEA-EOMCCSD(4$p$-2$h$)                    & 19.51\\
DIP-EOMCCSD(3$h$-1$p$)                    & 21.24\\
DIP-EOMCCSD(4$h$-2$p$)\{3\}\tnote{b}      & 19.31\\
DIP-EOMCCSD(4$h$-2$p$)                    & 19.31\\[1mm]
DEA-EOMCCSDt(4$p$-2$h$)\{3\}\tnote{a,c}   & 19.18\\
DIP-EOMCCSDt(4$h$-2$p$)\{3\}\tnote{b,d}   & 19.32\\[1mm]
DEA-EOMCCSDT(4$p$-2$h$)\{3\}\tnote{a}     & 19.22\\
DEA-EOMCCSDT(4$p$-2$h$)                   & 19.68\\
DIP-EOMCCSDT(4$h$-2$p$)\{3\}\tnote{b}     & 19.19\\
DIP-EOMCCSDT(4$h$-2$p$)                   & 19.19\\
Expt.\tnote{e}                            & $16.1 \pm 0.1$\\
${\rm Expt.} - \Delta{\rm ZPVE}$\tnote{f} & 18.1\\
\hline\hline
\end{tabular}
\end{minipage}

\begin{tablenotes}
\footnotesize
\item[a]{
The active orbitals used to select the dominant 4$p$-2$h$ excitations in the active-space DEA-EOMCC
calculations consisted of the three lowest-energy unoccupied orbitals of the $({\rm TMM})^{2+}$ reference system, which are the doubly degenerate $1e^{\prime\prime}$ and non-degenerate $2a_{2}^{\prime\prime}$ orbitals for the $D_{3h}$-symmetric $X \; {^{3}}A_{2}^{\prime}$ state and the $1a_{2}$, $2b_{1}$, and $3b_{1}$ orbitals for the $C_{2v}$-symmetric $B \; {^{1}}A_{1}$ state.
}
\item[b]{
The active orbitals used to select the dominant 4$h$-2$p$ excitations in the active-space DIP-EOMCC
calculations consisted of the three highest-energy occupied orbitals of the $({\rm TMM})^{2-}$ reference
system, which are the non-degenerate $1a_{2}^{\prime\prime}$ and doubly degenerate $1e^{\prime\prime}$ orbitals for the $D_{3h}$-symmetric $X \; {^{3}}A_{2}^{\prime}$ state and the $1b_{1}$, $1a_{2}$, and $2b_{1}$ orbitals
for the $C_{2v}$-symmetric $B \; {^{1}}A_{1}$ state.
}
\item[c]{
The active orbitals used to define the ground-state CCSDt calculations preceding the DEA-EOMCCSDt(4$p$-2$h$)$\{3\}$ diagonalizations consisted of the highest occupied non-degenerate $1a_{2}^{\prime\prime}$ orbital and the lowest unoccupied doubly degenerate $1e^{\prime\prime}$ and non-degenerate $2a_{2}^{\prime\prime}$ orbitals of $({\rm TMM})^{2+}$ for the
$D_{3h}$-symmetric $X \; {^{3}}A_{2}^{\prime}$ state and the highest occupied $1b_{1}$ orbital
and the lowest unoccupied $1a_{2}$, $2b_{1}$, and $3b_{1}$ orbitals of $({\rm TMM})^{2+}$ for
the $C_{2v}$-symmetric $B \; {^{1}}A_{1}$ state.
}
\item[d]{
The active orbitals used to define the ground-state CCSDt calculations preceding the DIP-EOMCCSDt(4$h$-2$p$)$\{3\}$ diagonalizations consisted of the highest occupied non-degenerate $1a_{2}^{\prime\prime}$ and doubly degenerate $1e^{\prime\prime}$ orbitals and the lowest unoccupied non-degenerate $2a_{2}^{\prime\prime}$ orbital of $({\rm TMM})^{2-}$ for the
$D_{3h}$-symmetric $X \; {^{3}}A_{2}^{\prime}$ state and the highest occupied
$1b_{1}$, $1a_{2}$, and $2b_{1}$ orbitals and the lowest unoccupied $3b_{1}$ orbital of
$({\rm TMM})^{2-}$ for the $C_{2v}$-symmetric $B \; {^{1}}A_{1}$ state.
}
\item[e]{
The experimentally determined adiabatic singlet--triplet separation taken from
Ref.\ \cite{TMM-Expt.}.
}
\item[f]{
The estimate of the purely electronic $\Delta E_{\rm S-T}$ gap obtained by subtracting
the zero-point vibrational energy correction, $\Delta{\rm ZPVE}$, determined using the
SF-DFT/6-31G(d) calculations in Ref.\ \cite{ch2_krylov}, from the
experimental singlet--triplet separation reported in Ref.\ \cite{TMM-Expt.}.
}
\end{tablenotes}
\end{threeparttable}
\end{table*}

\newpage
\clearpage
\pagebreak
\begin{table*}[h]
\centering
\footnotesize
\begin{threeparttable}
\caption{
The statistical errors characterizing the vertical DIPs, in eV, corresponding to the
lowest singlet and triplet states of the dicationic species obtained in the 
DIP-EOMCCSD(3$h$-1$p$), DIP-EOMCCSD(4$h$-2$p$), DIP-EOMCCSD(T)(a)(4$h$-2$p$),
DIP-EOMCCSDt(4$h$-2$p$)$\{N_{o}\}$, and DIP-EOMCCSDT(4$h$-2$p$) calculations for the
23 atoms and molecules considered in Refs. \cite{loos-ip,loos-ppbse} relative to the
high-level benchmark DIP values resulting from the CIPSI computations extrapolated to the
exact, full CI, limit reported in Ref. \cite{loos-ppbse}.
All calculations employed the aug-cc-pVTZ basis set \cite{ccpvnz,augccpvnz} and the lowest-lying orbitals correlating with the chemical cores of B, C, N, O, F, Ne, Si, S, Cl, and Ar were frozen
in all post-RHF steps. The DIP-EOMCC calculations used the RHF orbitals of the respective neutral
systems.
}
\label{23}
\begin{tabular}{lccccc}
\hline\hline
errors& 
\multicolumn{1}{c}{CCSD($3h\mbox{-}1p$)\tnote{a}}& 
\multicolumn{1}{c}{CCSD($4h\mbox{-}2p$)\tnote{b}}& 
\multicolumn{1}{c}{CCSD(T)(a)($4h\mbox{-}2p$)\tnote{c}}& 
\multicolumn{1}{c}{CCSDt($4h\mbox{-}2p$)\tnote{d}}& 
\multicolumn{1}{c}{CCSDT($4h\mbox{-}2p$)\tnote{e}} \\
\hline
\multicolumn{6}{c}{Singlets}\\
MSE\tnote{f} & 0.59 & $-$0.23    & 0.04 & $-$0.09 & 0.04  \\
MUE\tnote{g} & 0.59 & 0.23       & 0.07 & 0.10    & 0.05 \\
MaxE\tnote{h}& 1.20 & 0.67    & 0.43 & 0.18 & 0.23 \\
\multicolumn{6}{c}{Triplets}\\
MSE\tnote{f} & 0.64 & $-$0.24 & 0.04 & $-$0.09 & 0.04 \\
MUE\tnote{g} & 0.64 & 0.24    & 0.07 & 0.10    & 0.05 \\
MaxE\tnote{h}& 1.21 & 0.61 & 0.43 & 0.17 & 0.26 \\
\hline\hline
\end{tabular}

\begin{tablenotes}
\footnotesize
\item[a]{
The DIP-EOMCCSD(3$h$-1$p$) approach.
}
\item[b]{
The DIP-EOMCCSD(4$h$-2$p$) approach.
}
\item[c]{
The DIP-EOMCCSD(T)(a)(4$h$-2$p$) approach introduced in Ref.\ \cite{DIP-EOMCCSDT}.
}
\item[d]{
The DIP-EOMCCSDt(4$h$-2$p$)$\{N_{o}\}$ approach using small numbers of active unoccupied orbitals
($N_{u} \ll n_{u}$), listed in Table S3 of the Supplementary Data, in the underlying CCSDt calculations, 
while treating all correlated occupied orbitals as active (so that $N_{o}=n_{o}$).
}
\item[e]{
The full DIP-EOMCCSDT(4$h$-2$p$) approach.
}
\item[f]{
The mean signed error computed relative to the extrapolated CIPSI DIP values reported in Ref.\ \cite{loos-ppbse}.
}
\item[g]{
The mean unsigned error relative to the extrapolated CIPSI DIP values reported in Ref.\ \cite{loos-ppbse}.
}
\item[h]{
The maximum unsigned error relative to the extrapolated CIPSI DIP values reported in Ref.\ \cite{loos-ppbse}.
}
\end{tablenotes}
\end{threeparttable}
\end{table*}

\end{document}